\documentclass[aps,prb,twocolumn,amsmath,amssymb,superscriptaddress,showkeys,showpacs]{revtex4}
\usepackage{graphicx}
\usepackage{amssymb}
\usepackage{natbib}
\usepackage{calc}
\usepackage{SIunits}
\usepackage{textcomp}
\usepackage{color}
\usepackage{float}

\usepackage{amsmath}

\usepackage{hyperref}
\hypersetup{
	colorlinks,
	citecolor=blue,
	filecolor=blue,
	linkcolor=blue,
	urlcolor=blue,
	pdfdisplaydoctitle=true}

\begin{document}
\title{Preformed Cooper Pairs in a Triclinic Iron Pnictide Superconductor}
	
\author{Zezhong Li}
\thanks{These authors made equal contributions to this paper}
\affiliation{Beijing National Laboratory for Condensed Matter Physics, Institute of Physics, Chinese Academy of Sciences, Beijing 100190, China}
\affiliation{University of Chinese Academy of Sciences, Beijing 100049, China}
\author{Wenshan Hong}
\thanks{These authors made equal contributions to this paper}
\affiliation{Beijing National Laboratory for Condensed Matter Physics, Institute of Physics, Chinese Academy of Sciences, Beijing 100190, China}
\affiliation{University of Chinese Academy of Sciences, Beijing 100049, China}
\affiliation{International Center for Quantum Materials, School of Physics, Peking University, Beijing 100871, China}
\author{Honglin Zhou}
\affiliation{Beijing National Laboratory for Condensed Matter Physics, Institute of Physics, Chinese Academy of Sciences, Beijing 100190, China}
\affiliation{University of Chinese Academy of Sciences, Beijing 100049, China}
\author{Xiaoyan Ma}
\affiliation{Beijing National Laboratory for Condensed Matter Physics, Institute of Physics, Chinese Academy of Sciences, Beijing 100190, China}
\affiliation{University of Chinese Academy of Sciences, Beijing 100049, China}
\author{Uwe Stuhr}
\affiliation{Laboratory for Neutron Scattering and Imaging, Paul Scherrer Institut, CH-5232 Villigen PSI, Switzerland}
\author{Kaiyue Zeng}
\affiliation{Anhui Province Key Laboratory of Condensed Matter Physics at Extreme Conditions, High Magnetic Field Laboratory, Chinese Academy of Sciences, Hefei 230031, China}
\author{Long Ma}
\thanks{malong@hmfl.ac.cn}
\affiliation{Anhui Province Key Laboratory of Condensed Matter Physics at Extreme Conditions, High Magnetic Field Laboratory, Chinese Academy of Sciences, Hefei 230031, China}
\author{Ying Xiang}
\affiliation{National Laboratory of Solid State Microstructures and Department of Physics, Collaborative Innovation Center of Advanced Microstructures, Nanjing University, Nanjing 210093, China}
\author{Huan Yang}
\affiliation{National Laboratory of Solid State Microstructures and Department of Physics, Collaborative Innovation Center of Advanced Microstructures, Nanjing University, Nanjing 210093, China}
\author{Hai-Hu Wen}
\affiliation{National Laboratory of Solid State Microstructures and Department of Physics, Collaborative Innovation Center of Advanced Microstructures, Nanjing University, Nanjing 210093, China}
\author{Jiangping Hu}
\affiliation{Beijing National Laboratory for Condensed Matter Physics, Institute of Physics, Chinese Academy of Sciences, Beijing 100190, China}
\affiliation{University of Chinese Academy of Sciences, Beijing 100049, China}
\affiliation{Songshan Lake Materials Laboratory, Dongguan, Guangdong 523808, China}
\author{Shiliang Li}
\thanks{slli@iphy.ac.cn}
\affiliation{Beijing National Laboratory for Condensed Matter Physics, Institute of Physics, Chinese Academy of Sciences, Beijing 100190, China}
\affiliation{University of Chinese Academy of Sciences, Beijing 100049, China}
\affiliation{Songshan Lake Materials Laboratory, Dongguan, Guangdong 523808, China}
\author{Huiqian Luo}
\thanks{hqluo@iphy.ac.cn}
\affiliation{Beijing National Laboratory for Condensed Matter Physics, Institute of Physics, Chinese Academy of Sciences, Beijing 100190, China}
\affiliation{Songshan Lake Materials Laboratory, Dongguan, Guangdong 523808, China}
	
\date{\today}
	
\begin{abstract}
Electron pairing along with phase coherence generates superconductivity below the critical temperature ($T_c$). In underdoped high-$T_c$ cuprates, these two quantum phenomena may occur at separate temperatures, which was lately confirmed in the quasi-two-dimensional (quasi-2D) iron chalcogenide superconductors. Here, we report a systematic investigation on the pre-pairing behavior in a triclinic iron pnictide superconductor (Ca$_{0.85}$La$_{0.15}$)$_{10}$(Pt$_3$As$_8$)(Fe$_2$As$_2$)$_5$ with $T_c \approx $ 30 K, where the superconductivity is quasi-2D manifested by the Berezinskii-Kosterlitz-Thouless behaviors. Inelastic neutron scattering experiments unambiguously reveal a spin resonance peak around $E_R =$ 13 meV in the superconducting state, but its intensity continuously decreases when warming up across $T_c$, accompanied with an anomaly around $T^{*}\approx$ 45 K in spin correlations, and a suppression by an in-plane magnetic field persisting to the same temperature. Below $T^{*}$, a significant Nernst signal and a reduction of density of states at the Fermi level are also observed. These results suggest that the precursor of spin resonance is highly related to the preformed Cooper pairs driven by phase fluctuations, much like the pseudogap case in cuprates.
\end{abstract}


\maketitle
	
\section{Introduction}
Superconductivity arises from the long-range phase coherence and condensation of electron pairs below the critical temperature ($T_c$). In conventional superconductors described by Bardeen-Cooper-Schrieffer theory (BCS-theory), the phonon mediated Cooper pairs form and condense simultaneously at $T_c$ due to the large superfluid density \cite{schrieffer1999}. However, strong phase fluctuations can destroy the superconducting state and lead to a phase-incoherent state \cite{emery1995}, in which the Cooper pairs preform at a temperature $T^*$ above $T_c$. In superconducting thin films with disorders and low superfluid density, the superconducting phase transition is so-called as the Berezinskii-Kosterlitz-Thouless (BKT) type \cite{kosterlitz1973,kosterlitz1974}, where the zero resistance is broken  by phase fluctuations or equivalently by the unbinding of vortex-antivortex pairs. Within the BKT scenario, the current-voltage ($I-V$) characteristics of the material acquire a nonlinear dependence $V \propto I^{\alpha}$ near $T_c$, and the exponent $\alpha$ is larger than 3 below the BKT transition temperature ($T_{\mathrm{BKT}}$) \cite{minnhagen1995,medvedyeva2000,sharma2022}. In the phase incoherent state, the thermally generated vortices contribute to a Nernst signal \cite{spathis2008}, and the formation of incoherent Cooper pairs decreases in the density of states (DOS) below $T^*$ \cite{lascialfari2009,mondal2011}. Such a picture is adapted in high-$T_c$ cuprate superconductors, where the strong correlation induced low carrier density and the quasi-two-dimensional (quasi-2D) nature conspire to enhance the phase fluctuations \cite{uemura1989, corson1999, bollinger2011}. Indeed, a superconducting fluctuation regime above the superconducting dome has been identified in many cuprate systems by numerous Nernst measurements \cite{xu2000,yayu2001,yayu2002,yayu2003,yayu2006,ussishkin2002,kokanovic2009,chang2012,tafti2014,choiniere2018}, which has been taken as the direct evidence of preformed Cooper pairs. In addition, an incoherent pairing scenario has been put forward to explain the origin of the pseudogap state in the underdoped regime of cuprate superconductors, though it remains controversial. The pseudogap-like behaviors induced by phase fluctuations are also observed in those quasi-2D iron chalcogenide superconductors, such as the single-layer FeSe film on SrTiO$_3$ substrate and ion intercalated FeSe \cite{shaolong2013,shiyong2013,whzhang2014,yxu2021,faeth2021,kang20201,Kang20222,dli2022}. Some spectroscopic measurements on iron pnictide superconductors also give possible evidence of the pseudogap state, but its microscopic origin is still not clearly understood \cite{ymxu2011,shbaek2011,khlin2014,sjmoon2014,tshimojima2014,hyang2016,czhang2022,sshao2023,mcli2022}.

In unconventional superconductors, instead of the phonon excitations acting as the pairing glue in BCS superconductors, spin fluctuations are arguably the common thread in understanding the pairing and condensation of Cooper pairs \cite{scalapino2012,eschrig2006,tranquada2014,white2015,paglione2010,pdai2012,pdai2015,xchen2014,si2016,dgong2018,rxliu2023,gu2022,zjliu2023,wwu2024,mwang2024,qqin2024}. In most optimally doped cuprate superconductors, a collective exciton known as the neutron spin resonance mode (SRM) occurs below $T_c$, characterized by a sharp peak in the spin excitations around the antiferromagnetic wave vector $Q_{\mathrm{AF}}$ \cite{eschrig2006,tranquada2014}. The temperature dependence of its intensity behaves like a superconducting order parameter \cite{mignod1991,sidis2007,gyu2009}. Whereas, in the underdoped YBa$_2$Cu$_3$O$_{6+\delta}$ and HgBa$_2$CuO$_{4+\delta}$, it has been observed that the resonance-like spin fluctuations at the resonance energy $E_R$ persist up to a doping-dependent temperature $T^*$, which is consistent with the pseudogap temperature $T_p$ determined by other measurements \cite{pdai1999,pdai2001,hinkov2007,chan2016a,chan2016b,hashimoto2014}. Furthermore, these fluctuations exhibit the same anisotropic magnetic field suppression effect as in the superconducting state, supporting an incoherent pairing scenario in the pseudogap state above $T_c$ \cite{pdai2000}. The existence of SRM in iron-based superconductors (FeSCs), including iron chalcogenides and iron pnictides, has been extensively revealed through inelastic neutron scattering (INS) experiments \cite{christianson2008,qiu2009,tjliu2010,inosov2010,park2010,mwang2010,czhang2011,hqluo2012,clzhang2013,chlee2013,hqluo2013,jzhao2013,qwang2016,dhu2016,mma2017,txie2018a,txie2018b,rzhang2018,txie2021,whong2020,cliu2022,mma2023,whong2023,zzli2024}. However, it is not clear whether the incoherent Cooper pairs induced by phase fluctuations can be traced by the residual intensity of the SRM above $T_c$ in FeSCs.

Among the FeSCs, a triclinic iron pnictide superconductor family Ca$_{10}$(Pt$_3$As$_8$)(Fe$_2$As$_2$)$_5$ (10-3-8) exhibits a high anisotropy \cite{xiang2012,neupane2012} and a low superfluid density \cite{kim2012,cho2012,seo2018}, providing a possible platform to search the preformed Cooper pairs induced by phase fluctuations \cite{qywu2023,ryang2019}. Infrared reflectivity spectra on the optimally doped La and Na in the 10-3-8 compound have revealed that the low-frequency Drude spectrum develops a pseudogap hump structure \cite{ryang2019,seo2017,choi2021}. In the optimally Pt-doped (CaFe$_{1-x}$Pt$_x$As)$_{10}$Pt$_3$As$_8$ (CaPt-10-3-8) with $T_c \approx$ 13 K, the spin lattice relaxation rate divided by temperature 1/$T_1T$ shows a drop below $T^*$ = 45 K, implying a loss of DOS above $T_c$. Meanwhile, INS studies on the CaPt-10-3-8 reveal that a peak around $E =$ 7 meV appears below $T^*$. However, this mode cannot be directly associated with the SRM, as there is no intensity gain or gapped features in the spin excitations detected below $T_c$ \cite{surmach2015}. In the (CaFe$_{1-x}$Pt$_x$As)$_{10}$Pt$_4$As$_8$ (CaPt-10-4-8) compounds with $T_c$ $\approx$ 30 K, a peak at 12 meV in $\chi^{\prime\prime}(Q,\omega)$ emerges in the superconducting state, but it may be mixed by phonon excitations and its temperature dependence does not behave like an order parameter when crossing $T_c$ \cite{sato2011,kikeuchi2014,kikeuchi2015}. 	
	
\begin{figure}[htbp]
	\includegraphics[width=0.45\textwidth]{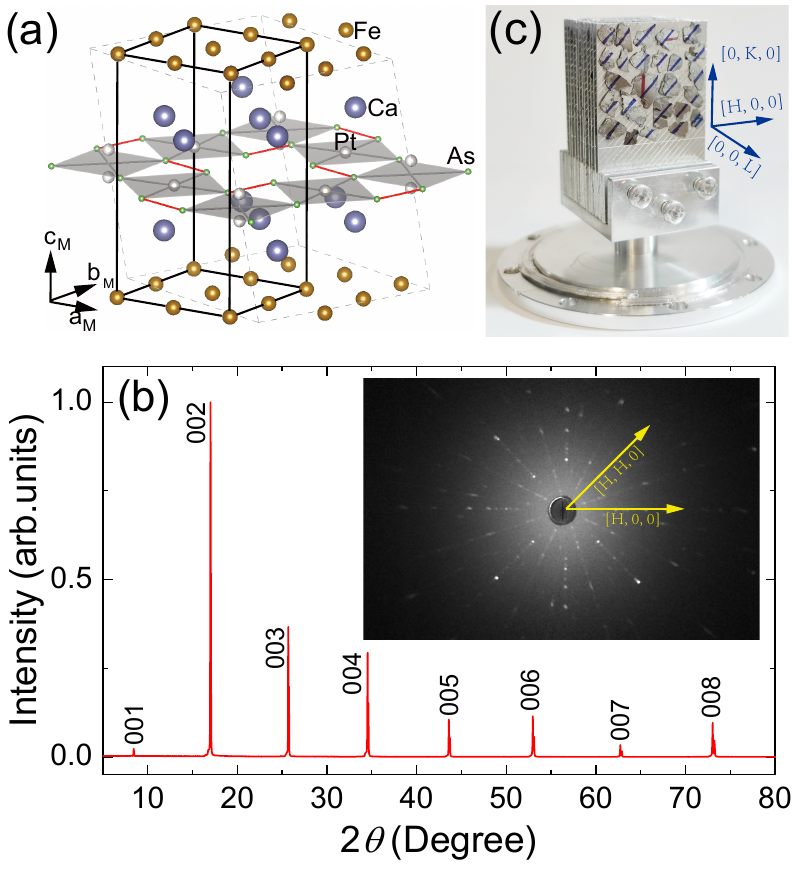}
	\caption{(Color online) Crystal structure and characterization of CaLa-10-3-8 samples.
		(a) Crystalline structure of CaLa-10-3-8. The black box represents the magnetic unit cell defined by Fe sublattice.
		(b) Photographs of assembled and co-aligned CaLa-10-3-8 crystals in our INS experiments.
		(c) X-ray diffraction pattern of a typical CaLa-10-3-8 crystal. The inset shows the Laue reflection pattern of the crystal.
	}
\label{fig1}
\end{figure}

In the present work, based on the high quality single crystals of optimally La-doped (Ca$_{1-x}$La$_{x}$)$_{10}$(Pt$_3$As$_8$)(Fe$_2$As$_2$)$_5$ ($x=0.15$) (CaLa-10-3-8) with quasi-2D superconductivity below $T_c \approx$ 30 K, we systematically explore the low-energy spin excitations, electrical transport, Nernst effect, and nuclear magnetic resonance (NMR) spectroscopy. We present concrete evidence for a quasi-2D SRM at $E_R=$ 13 meV by INS. However, when warming up across $T_c$, the susceptibility $\chi^{\prime\prime}(Q,\omega)$ at $E_R$ continuously decreases. Both the spin correlations and $\mathrm{d}\chi^{\prime\prime}(Q,\omega)/\mathrm{d}T$ exhibit a clear anomaly around $T^*$ = 45 K. The spin excitation intensity can be suppressed by an in-plane magnetic field just below $T^{*}$. Furthermore, a significant field-dependent Nernst signal and a reduction of 1/$T_1T$ in NMR are also observed below $T^{*}$. These phenomena exhibit similarities with the cases in the pseudogap state of underdoped cuprates \cite{pdai1999,pdai2001,hinkov2007,chan2016a,chan2016b,hashimoto2014}, suggesting the preformed Cooper pairs induced by phase fluctuations above $T_c$. Therefore, the pre-pairing behavior universally exists in those quasi-2D unconventional superconductors and likely relates to the spin-spin interactions.

\section{Experimental methods}
The (Ca$_{0.85}$La$_{0.15}$)$_{10}$(Pt$_3$As$_8$)(Fe$_2$As$_2$)$_5$ single crystals were grown with the self-flux method according to previous reports \cite{nni2011,nni2013}. The CaLa-10-3-8 crystallizes in a triclinic structure (space group P-1) due to the intermediary Pt$_3$As$_8$ layers, while the magnetic unit cell of Fe sublattice keeps nearly orthorhombic as shown in Fig. \ref{fig1}(a). The sharp $(0, 0, L)$ peaks in Fig. \ref{fig1}(b) indicate high $c$-axis orientation of our sample. Our INS experiments were carried out using two thermal triple-axis spectrometers: EIGER at the Swiss Spallation Neutron Source (SINQ), Paul Scherrer Institute \cite{stuhr2017}, and BAMBOO at China Advanced Research Reactor (CARR), China Institute of Atomic Energy. The scattering plane $[H, 0, 0]$ $\times$ $[0, 0, L]$ is defined by $Q$ = ($H$, $K$, $L$) = ($q_xa/2$$\pi$, $q_yb/2$$\pi$, $q_zc/2z$$\pi$) in reciprocal lattice units (r.l.u.) using a pseudo-orthorhombic magnetic unit cell with $a_M \approx b_M$ $\approx$ 5.54 ${\rm \AA}$, $c_M$ = 10.27 ${\rm \AA}$. We co-aligned about 8.3 g ($\sim$580 pieces) of single crystals on several aluminum plates using an X-ray Laue camera. The inset of Fig. \ref{fig1}(b) shows the Laue reflection pattern of a CaLa-10-3-8 single crystal, where $[H, 0, 0]$ and $[H, H, 0]$ directions are marked. The assembled crystal set is shown in Fig. \ref{fig1}(c) with mosaic about 3$\degree$ for both scattering axes. INS data were collected with a final energy fixed as $E_f$ = 14.7 meV in energy loss measurement mode ($E_i$ $\textgreater$ $E_f$), with a pyrolytic graphite filter, a double-focusing monochromator, and a vertical focusing analyzer in setup.

\begin{figure}[htbp]
	\includegraphics[width=0.48\textwidth]{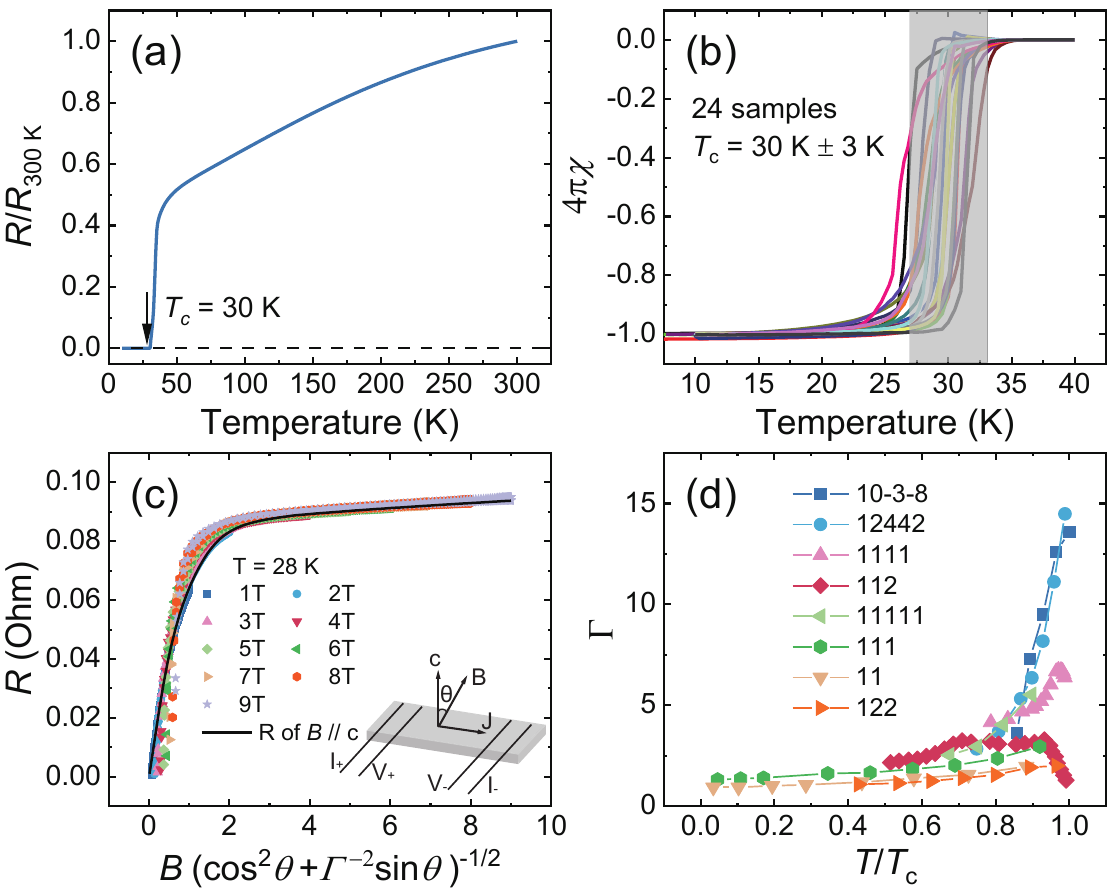}
	\caption{(Color online) Resistance and magnetization characterization of CaLa-10-3-8 crystals.
		(a) Temperature dependence of the resistance within $ab-$plane. The superconducting transition at $T_c$ = 30 K is marked by a arrow.
		(b) Zero-field-cooling magnetization in 24 randomly selected crystals. The statistic on superconducting transition is about $T_c$ = 30 $\pm$ 3 K.
		(c) Scaling of the resistivity versus an effective field, $B(\rm{cos}^2(\theta)+\mathit{\Gamma}^{-2}\rm{sin} (\theta))^{-1/2}$, where $\mathit{\Gamma}$ represents the superconducting anisotropy defined by the upper critical field: $\mathit{\Gamma}$ = $B^{ab}_{c2}/B^c_{c2}$. The inset shows the schematic configuration of the magnetic
		resistance measurements.
		(d) Comparison of the superconducting anisotropy $\mathit{\Gamma}$ for various FeSCs.
	}
\label{fig2}
\end{figure}

The quality of CaLa-10-3-8 crystals was examined by the resistance and magnetization measurements as shown in Fig \ref{fig2}(a) and (b). The electric transport and magnetic-susceptibility measurements were carried out on the physical property measurement systems (PPMS) and magnetic property measurement systems (MPMS) (Quantum Design), respectively. The sharp superconducting transitions for 24 randomly selected samples indicate the homogeneity and high quality of our samples, with a statistic $T_c$ = 30 $\pm$ 3 K. We also measured the in-plane magneto-resistance within the superconducting transition around $T$ = 28 K to characterize the superconducting anisotropy. According to the anisotropic Ginzburg-Landau theory, the resistivity in the mixed state can be scaled with the variable $B\sqrt{\rm{cos}^2(\theta)+\mathit{\Gamma}^{-2}\rm{sin}^2(\theta)}$, where the angle $\theta$ is between the $c_M$-axis and magnetic field, $\mathit{\Gamma} = B^{ab}_{c2}/B^{c}_{c2}$ is the superconducting anisotropy. Fig. \ref{fig2}(c) shows the scaling results of magnetoresistance at $T$ = 28 K, which give $\mathit{\Gamma}$ = 13.6. Comparing the superconducting anisotropy between the CaLa-10-3-8 and other FeSCs after normalizing by $T_c$ as shown in Fig. \ref{fig2}(d) \cite{twang2020}, the CaLa-10-3-8 and 12442 system show the largest anisotropy $\mathit{\Gamma}$ $\textgreater$ 10 indicating their quasi-2D nature.
	
The Nernst effect was measured in a homemade setup attached to PPMS. A temperature gradient was established in the $ab-$plane of the sample by a heater, and the temperature difference was measured by a pair of type-E thermocouples. The Nernst signals $S_{xy}$ are measured by the transverse voltage ($V$) produced from a longitudinal temperature gradient (-$\triangledown T$) in a perpendicular magnetic field ($B$ $\parallel$ $c$).
	
The NMR measurements were conducted at the $^{75}$As nuclear sites with a phase-coherent spectrometer. There are two very different $^{75}$As environments including the FeAs$_4$ tetrahedra and the intercalated Pt$_3$As$_8$ skutterudite layer. However, only one set of resonance peaks corresponds to the $^{75}$As nuclear spins nearby the Fe ions, identical to the previous NMR study on the undoped sample \cite{zhou2013}. The linewidth of the resonance peak is very broad (0.31 MHz at 40 K) compared to the results of other FeSCs due to the existing five magnetically inequivalent As sites in the FeAs layers and the disorder resulting from the Pt impurities. But the smaller relative width $\Delta f/f$ = 0.28$\%$ than previous reports in 10-3-8 system \cite{zhou2013, surmach2015} suggests that our crystals have less Pt impurities. The measurement of the nuclear spin-lattice relaxation rate (1/$T_1$) was performed by the inversion-recovery method. The 1/$T_1$ was obtained by fitting the time evolution of nuclear magnetization of the central transition to the function:
\begin{equation}
	\begin{aligned}
		M(t)/M(\infty) &= \\
		1 - a(0.1&\exp(-(t/T_1)^\beta) + 0.9\exp(-(6t/T_1)^\beta))
	\end{aligned}
\end{equation}
The fitting results give a stretching factor $\beta$ $\textgreater$ 0.7, suggesting a slight distribution of 1/$T_1$ across the sample.

\section{Results}

\begin{figure}[htbp] \centering
	\includegraphics[width=0.4\textwidth]{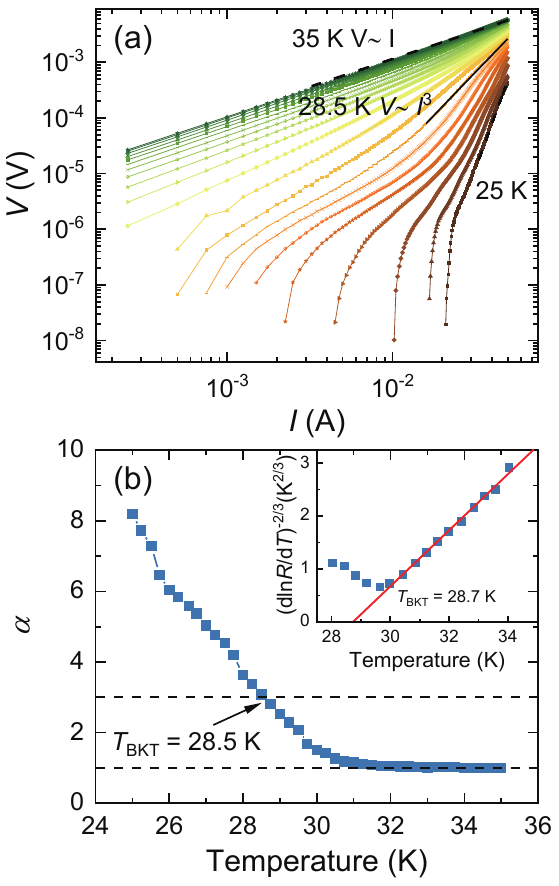}
	\caption{(Color online) BKT transition analysis on the $I-V$ curves of CaLa-10-3-8 crystals.
		(a) Raw data of $I-V$ curves at various temperatures plotted in a logarithmic scale.
		(b) Temperature dependence of the power-law exponent $\alpha$ as deduced from $V \sim I^{\alpha} $. The inset shows the temperature dependence of (dln$R$/d$T$)$^{-2/3}$ and the red solid line shows the fitting by the Halprin-Nelson formula \cite{halperin1979}.
	}
	\label{fig3}
\end{figure}

\begin{figure*}[htbp]	\centering
	\includegraphics[width=0.9\textwidth]{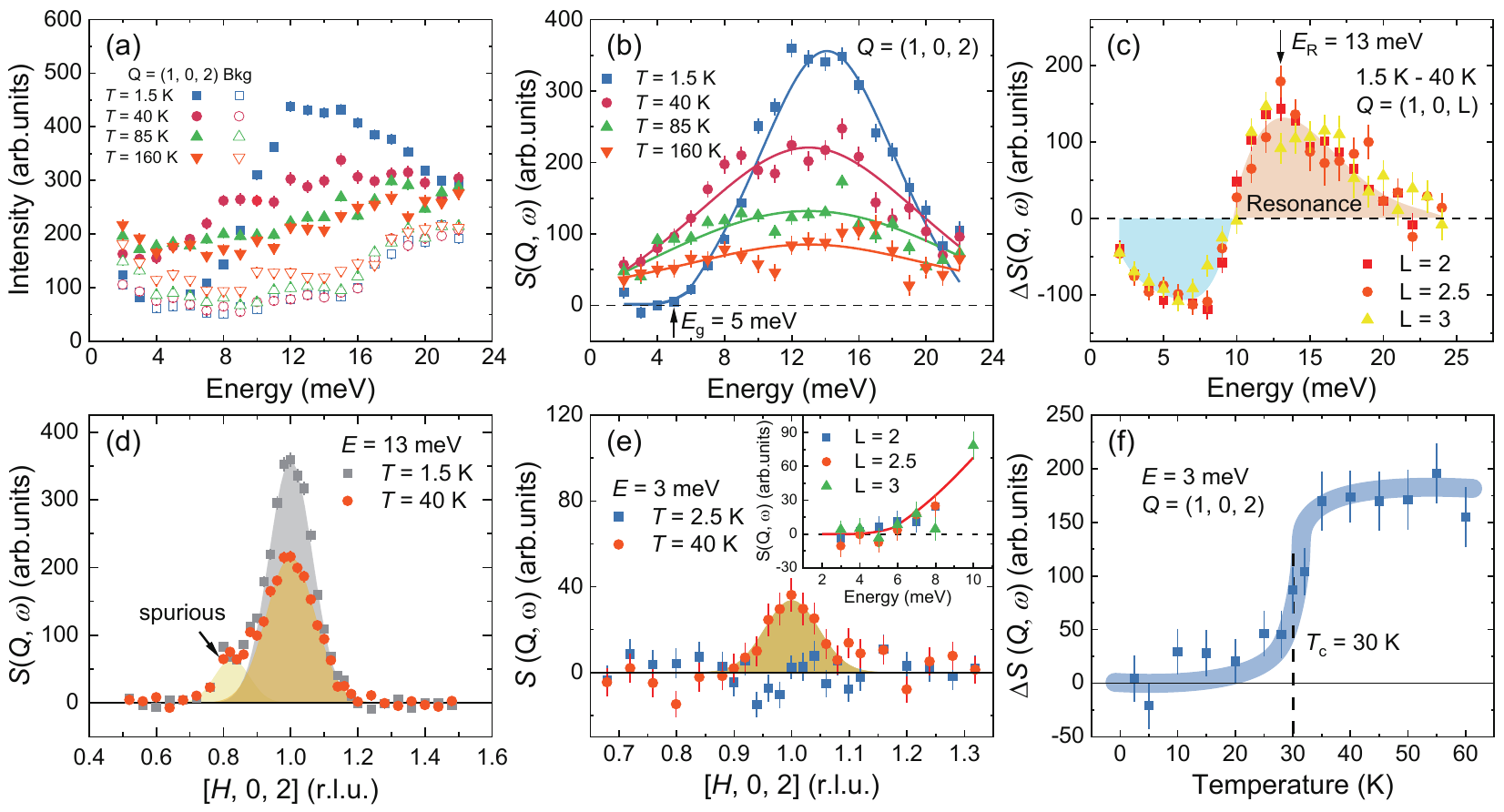}
	\caption{(Color online) Low-energy spin excitations of CaLa-10-3-8 measured by INS.
		(a) Raw data of energy scans for spin excitations at $Q$ = (1, 0, 2) at $T$ = 1.5 K, 40 K, 85 K and 160 K measured at EIGER. The open shape in (c) shows the background at $Q$ = (1.3, 0, 2) for E = 2 - 12 meV and $Q$ = (1.5, 0, 2) for E = 13 - 22 meV.
		(b) Energy dependence of spin excitations at $Q=$ (1, 0, 2), $T$ = 1.5, 40, 85 and 160 K.
		(c) The intensity difference between $T$ = 1.5 and 40 K at $Q = (1, 0, L)$ ($L =$ 2, 2.5, 3). All data are normalized by the square of Fe$^{2+}$ magnetic form factor $|F(Q)|^2$.
		(d) $Q$-scans along $[H, 0, 2]$ for $E$ = 13 meV at $T=$ 1.5 and 40 K measured at EIGER. The light yellow region marks an spurious peak from the harmonic scattering.
		(e) $Q$-scans along $[H, 0, 2]$ for $E$ = 3 meV at $T=$ 2.5 and 40 K measured at BAMBOO. The inset shows the spin gap at different $L$s.
		(f) Temperature dependence of the spin gap intensity at $E=$ 3 meV, $Q=$(1, 0, 2) measured at BAMBOO.  All solids lines are guides to eyes.
	}
\label{fig4}
\end{figure*}	
\subsection{Electronic transport}
The incoherent Cooper pairs are usually associated with strong phase fluctuations approaching the 2D limit of superconductivity in which the transition can be described under the BKT framework. Below the BKT-transition, the Ohm's law is broken due to the unbinding of vortex-antivortex pairs, resulting in a nonlinear behavior in the current-voltage ($I-V$) curves. We first present the temperature-dependent $I-V$ curves for a CaLa-10-3-8 sample with $T_c$ about 29 K to verify whether the superconducting transition is of the BKT type. We extracted the power-law exponent $\alpha$ for $V \sim I^\alpha$ by fitting the $I-V$ curves shown in Fig. \ref{fig3}(a). As the temperature decreases, the exponent $\alpha$ gradually deviates from an Ohmic resistance behavior and becomes highly temperature dependent. The $\alpha$ continuously increases to 3 at about 28.5 K, as shown in Fig. \ref{fig3}(b), which is a typical signature of the BKT transition in 2D superconductors \cite{minnhagen1995,medvedyeva2000,sharma2022,reyren2007}. Furthermore, the disappearance of Ohmic resistance follows the Halperin-Nelson formula \cite{halperin1979}:
\begin{equation}
R(T) = R_0\exp[-b/(T-T_\mathrm{BKT})^{1/2}]
\end{equation}
As shown in the plot of $[d(\mathrm{ln} R)/dT]^{2/3}$ [inset of Fig. \ref{fig3}(b)], the extracted value of $T_\mathrm{BKT}$ from the fitting is about 28.7 K, which is in agreement with the result analysis on the $I-V$ curves. For comparison, similar results were also observed in the FeSe/SrTiO$_3$ thin films and the ion intercalated FeSe materials \cite{kang20201,faeth2021}. Therefore, the $I-V$ curves and the temperature dependent resistance give unambiguous evidence of a BKT-like transition in the CaLa-10-3-8 system, confirming its quasi-2D nature of superconductivity shown in Fig. \ref{fig2}(c) and (d).

\subsection{Neutron spin resonance mode}
\begin{figure}[htbp] \centering
	\includegraphics[width=0.4\textwidth]{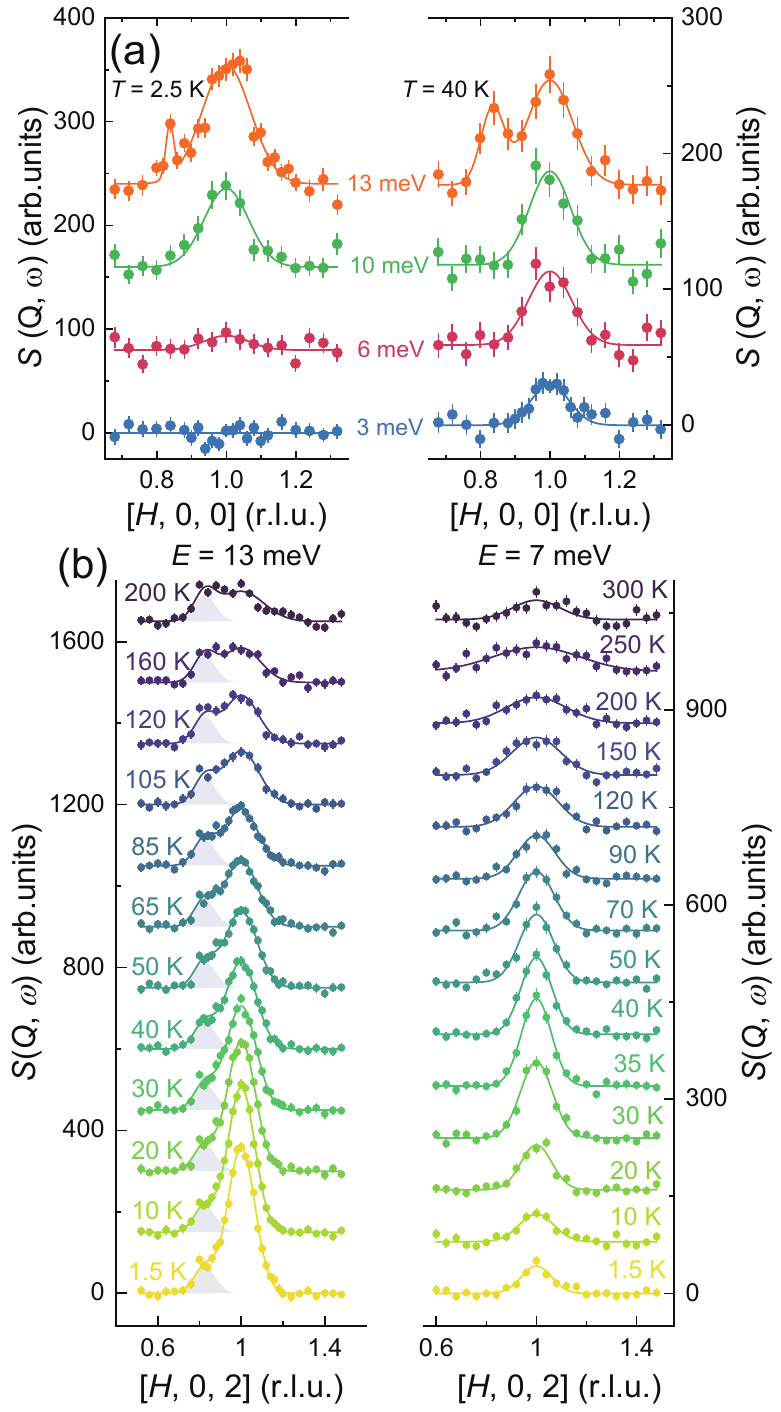}
	\caption{(Color online) Temperature dependence of the spin excitations of CaLa-10-3-8 measured by INS.
		(a) Constant-energy scans along $H-$direction for $E$ = 3, 6, 10 and 13 meV at $T$ = 2.5 K and $T$ = 40 K measured at BAMBOO. A linear background has been subtracted, and the solid lines are Gaussian fits by one or two peaks including the spurious signals around $Q=(0.8, 0, 2)$ for $E$ = 13 meV.
		(b) $Q$-scans along $Q=[H, 0, 2]$ for $E=13$ and $7$ meV at different temperatures measured at EIGER. The solid lines are fitting results by a Gaussian function, where a temperature-independent Gaussian peak (shadow areas) from the spurious signals around $Q=(0.8, 0, 2)$ is added to the fittings for $E = 13$ meV.
	}
\label{fig5}
\end{figure}

We then present the results on the spin excitations in Fig. \ref{fig4}. By subtracting the background as shown in Fig. \ref{fig4} (a), we obtain the energy dependence of spin excitation intensity S($Q$, $\omega$) at the antiferromagnetic wave vector $Q_{\rm AF}$=(1, 0, 2) as shown in Fig. \ref{fig4}(b). Upon entering into the superconducting state, a spin gap opens below 5 meV, and a broad peak centered at $E$ = 13 meV is observed. In the normal state, the peak transforms into a hump above $T_c$ with the intensity decreasing upon warming, similar to the results in CaPt-10-4-8 compound \cite{sato2011}. By comparing the spin excitation intensity below and above $T_c$ [Fig. \ref{fig4}(c)], a SRM with peak energy $E_R=$ 13 meV is identified by a clear intensity enhancement from 10 meV to 20 meV and a depletion below 10 meV, inconsistent with the previous results of CaPt-10-3-8 \cite{surmach2015}. Furthermore, all data for different $L$ = 2, 2.5, and 3 overlap after being modified by the magnetic form factor $\lvert F(Q) \rvert ^2$ of Fe$^{2+}$ ions, suggesting the quasi-2D nature of the SRM. An enhancement of the intensity at $T=$ 1.5 K can also be detected around $Q=(1, 0, 2)$ in the constant-energy scan ($Q$-scan) at $E=$ 13 meV in comparison to the 40 K data, where a temperature-independent spurious peak at $Q=(0.8, 0, 2)$ is probably from the accidental harmonic scattering of the (0, 0, 6) nuclear peak [Fig. \ref{fig4}(d)]. The resonance energy $E_R=13$ meV yields $E_R/k_BT_c=5.0$, following the same scaling obtained in other FeSCs \cite{txie2018a,whong2020,zzli2024}. To confirm the spin gap, we also performed $Q$-scans at 3 meV. As shown in Fig. \ref{fig4}(e), no peak is detected at $T=$ 2.5 K, but a peak emerges centered around $Q=$ (1, 0, 2) at $T=$ 40 K, suggesting fully gapped spin excitations at low energies in the superconducting state. The spin gap is also quasi-2D in reciprocal space for its $L$-independent feature [inset of Fig. \ref{fig4}(e)], which emerges just below $T_c$ [Fig. \ref{fig4}(f)]. To compare the dispersions, we show the data of several constant-energy scans ($H-$scans) at $T=$ 2.5 K and 40 K and $E$ = 3, 6, 10, 13 meV in Fig. \ref{fig5}(a). The peak width continuously increases upon increasing energy with similar tendency in both the superconducting and normal state. Neither the hourglass shape of the dispersions in the superconducting state nor the chimney-like dispersions in the normal state typically shown in the hole underdoped cuprates are observed in this sample.

\begin{figure}[htbp] \centering
	\includegraphics[width=0.48\textwidth]{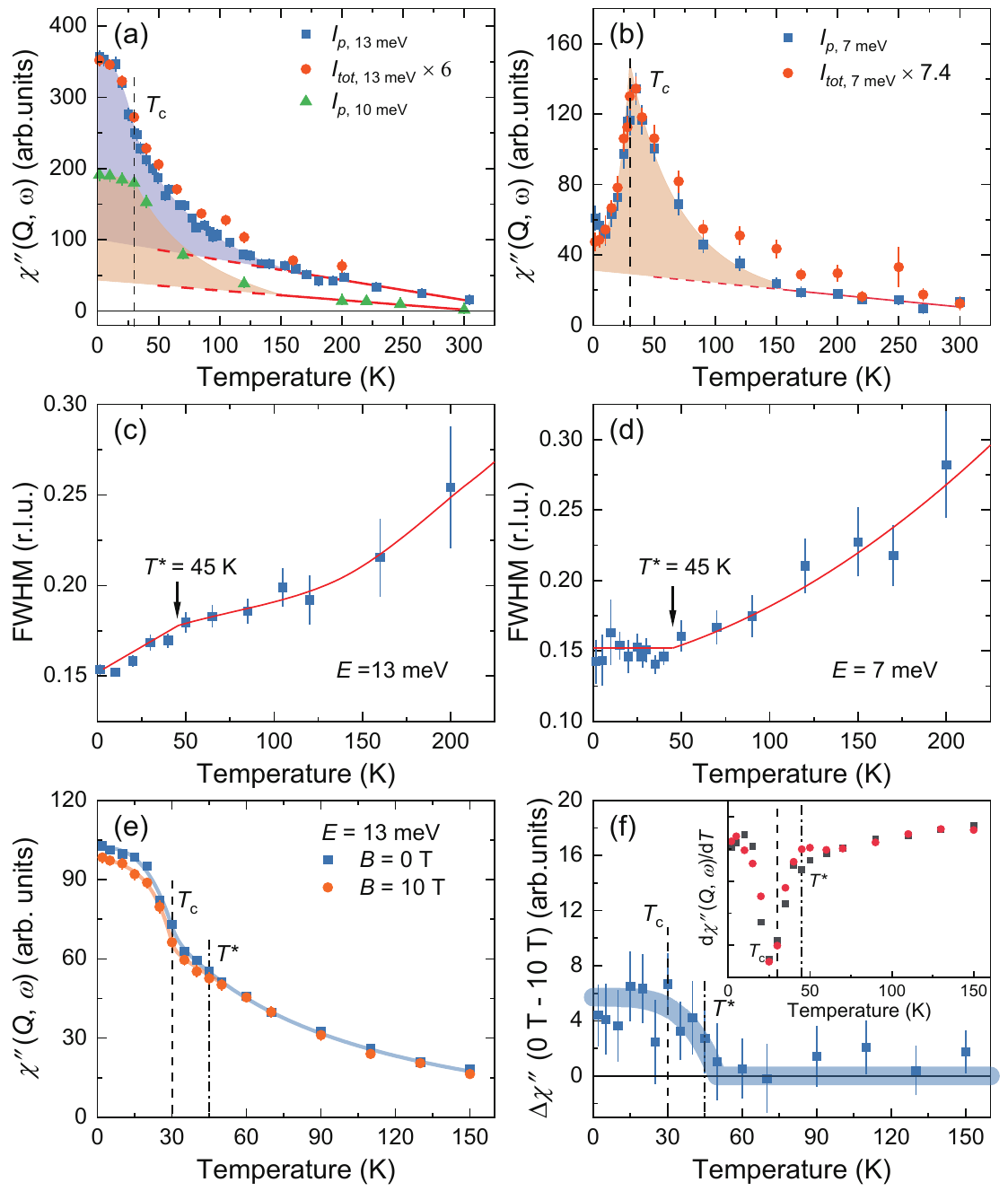}
	\caption{(Color online) Analysis on the temperature and magnetic field dependence of the spin excitations in CaLa-10-3-8.
		(a) and (b) Temperature dependence of $\chi^{\prime\prime} (Q,\ \omega)$ for $E$ = 13, 10 and 7 meV. Here $I_p$ and $I_{tot}$ imply the peak intensity and the integrated intensity around $Q =$ (1, 0, 2), respectively.
		(c) and (d) Temperature dependence of the peak width (FWHM) obtained by gaussian fittings to the constant-energy scans in Fig. \ref{fig5}(b). The solid lines are guides by the eyes, and the arrows mark the kinks of FWHM at $T^*=$ 45 K.
		(e) Temperature dependence of $\chi^{\prime\prime} (Q,\ \omega)$ for $E$ = 13 meV at 0 and 10 T.
		(f) Difference of $\Delta\chi^{\prime\prime} (Q,\omega) $ between 0 and 10 T. The inset shows the temperature derivative of $\chi^{\prime\prime} (Q,\ \omega)$, where both $T_c$ and $T^*$ are marked by dashed lines.
	}
\label{fig6}
\end{figure}

We have tracked the temperature evolution of the spin excitations both at the resonance energy $E=$ 13 meV and at $E=$ 7 meV in the depletion part. The $Q$-scans from $T=$ 1.5 to 300 K and their Gaussian fittings are shown in Fig. \ref{fig5}(b), where the spurious signal is a constant peak marked by the shadows. The integrated intensity ($I_{tot}$) and the full width at half maximum (FWHM) are obtained and presented in Fig. \ref{fig6}(a)-(d). To double-check the temperature dependence of magnetic intensities, we also measured the peak intensity ($I_p$) at $E=$ 10 meV and $E=$ 13 meV. All backgrounds are subtracted from the raw data, and the results are converted to dynamic magnetic susceptibility $\chi^{\prime\prime} (Q,\omega)$ by removing the Bose population factor. On warming up to $T_c$, the intensity at $E_R$ = 13 meV drops quickly due to the fading resonance. The intensity decrease does not cease at $T_c$ but continues to high temperature in the normal state [Fig. \ref{fig6}(a)]. The spin fluctuations usually decrease linearly at a high temperature due to the damping effect [see solid lines in Fig. \ref{fig6}(a)] \cite{hinkov2007,bourges2000}, thus the extra magnetic excitation at $E_R$ may be the precursor of SRM. A similar feature can be found in the excitation at $E=$ 10 meV. For the intensity at $E=$ 7 meV, it increases firstly due to the spectral weight redistribution and forms a $\lambda$ shape when across $T_c$, then decreases in the normal state, too [Fig. 4(b)]. Through the temperature dependence of FWHM, which is inversely associated with the spin-spin correlation lengths in real space by a Fourier transformation \cite{pdai2015}, we have found the FWHM monotonously increases upon warming up to room temperature with a clear kink around $T^{*}\approx$ 45 K, both for $E=$ 7 meV and $E=$ 13 meV [Fig. \ref{fig6}(c) and (d)].
	
To investigate the relationship between spin fluctuations above $T_c$ and superconductivity, we have also carried out magnetic field dependent measurements on the SRM in CaLa-10-3-8. Generally, a magnetic field will yield net changes of the magnetic intensity due to field-insensitive phonon excitations and incoherent scattering backgrounds. In the quasi-2D unconventional superconductors, a $c-$axis aligned field is expected to lower the resonance energy $E_R$ due to the quick suppression of $T_c$, while an in-plane magnetic field could effectively suppress the resonance intensity with limited effects on $E_R$ or $T_c$ \cite{tranquada2004,stephen2007,sli2011,jzhao2010}. As shown in Fig. \ref{fig6} (e) and (f), a 10 T in-plane magnetic field suppresses the intensity of spin fluctuations at the resonance energy $E_R$ below $T^*$, instead of the field effect below $T_c$ in other FeSCs \cite{sli2011,jzhao2010,ysong2018}. The inset in Fig. \ref{fig6} (f) depicts the temperature derivative of the data in Fig. \ref{fig6} (e), revealing a clear anomaly around $T^*$, associated with a rapid enhancement of spin fluctuations below $T_c$. These phenomena show similarities with the spin excitations in the underdoped cuprate superconductors \cite{pdai2000}. Therefore, the appearance of field-dependent spin fluctuations in between $T_c$ and $T^*$ represents a precursor resonant mode related to the formation of incoherent Cooper pairs, if we believe that the spin fluctuation indeed acts as the pairing glue in FeSCs.
	
\subsection{Nernst measurements}
\begin{figure}[htbp] \centering
	\includegraphics[width=0.48\textwidth]{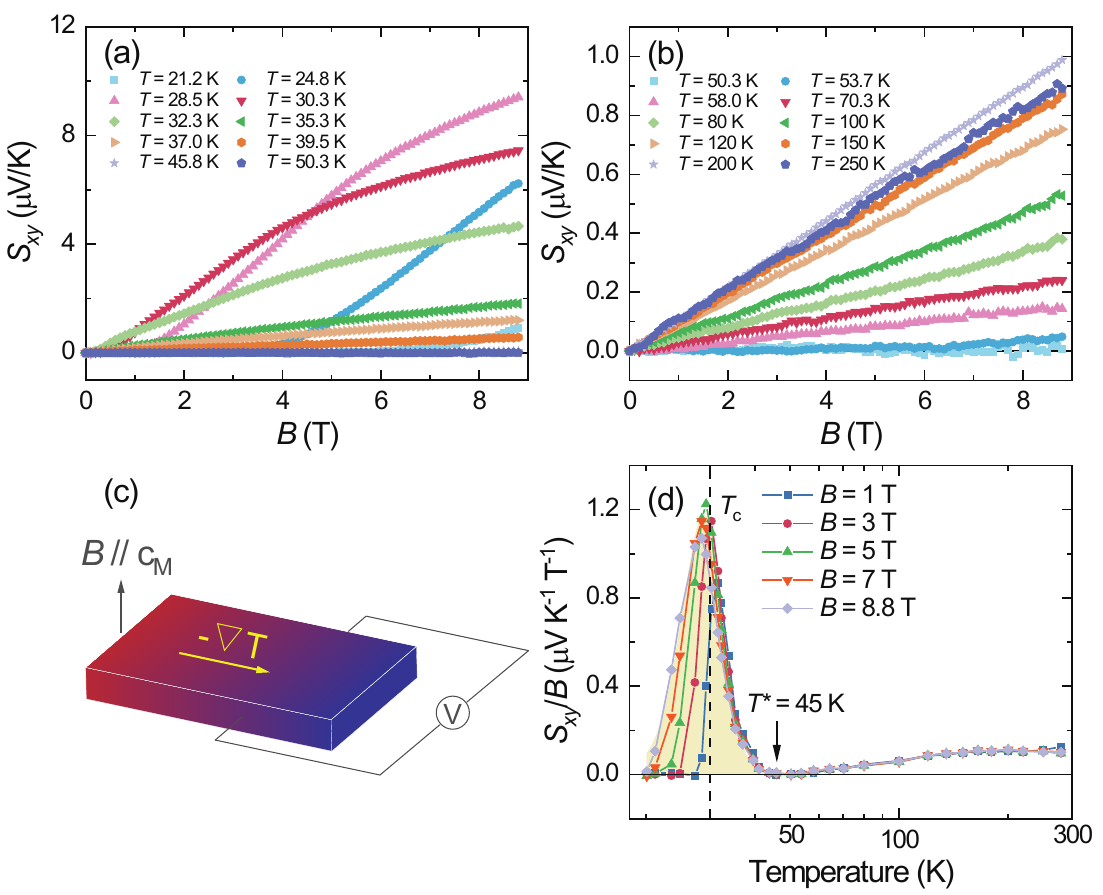}
	\caption{(Color online) Nernst measurement results on CaLa-10-3-8 crystals.
		(a) and (b) The field dependence of the Nernst signals at different temperatures.
		(c) Schematic of the Nernst measurement configuration.
		(d) Temperature dependence of the Nernst signals with an onset temperature of phase fluctuations at $T^*=$ 45 K.
	}
\label{fig7}
\end{figure}

Strong superconducting phase fluctuation provokes the unbinding of vortex-antivortex pairs in the normal state of the quasi-2D superconductors. Thus a transverse Nernst signal, generated by the vortex motion under a longitudinal temperature gradient and a vertical magnetic field, can be detected above $T_c$, which is usually taken as the hallmark of phase fluctuations in type-II superconductors \cite{kang20201,mcli2022,kbehnia2016}. We have performed the Nernst measurements on CaLa-10-3-8, and the results are shown in Fig. \ref{fig7}. Fig. \ref{fig7} (a) and (b) show the magnetic field dependence of measured transverse voltage under a longitudinal temperature gradient. A positive and nonlinear-field-dependent Nernst signal in the superconducting state is a feature of the vortex flow and vanishes at a lower temperature due to the vortex liquid to solid transition (Fig. \ref{fig7} (a)) \cite{yayu2006,choiniere2018,ssundar2010}. Intriguingly, the Nernst signal is also nonlinear above $T_c$ and persists up to $T^*$, which is shown more clearly in the temperature dependence of the deduced Nernst coefficient for different fields [Fig. \ref{fig6}(d)]. Since the Gaussian fluctuations can not account for such temperature-dependent behavior of the Nernst signal, they must be contributed by the superconducting phase fluctuation above $T_c$ \cite{ussishkin2002,tafti2014}. After warming up above $T^*$, the Nernst signals become linear with the magnetic field (Fig. \ref{fig7} (b)). The nonzero magnitude as well as the slow hump of $S_{xy}/B$ from $T=50$ K to 300 K are probably related to the multi-band characteristic of FeSCs, which were also observed in another quasi-2D FeSC compound CsCa$_2$Fe$_4$As$_4$F$_2$ but interpreted as an effect from incoherent-coherent crossover of charge dynamics \cite{mcli2022}.

\subsection{NMR measurements}
\begin{figure}[htbp] \centering
	\includegraphics[width=0.48\textwidth]{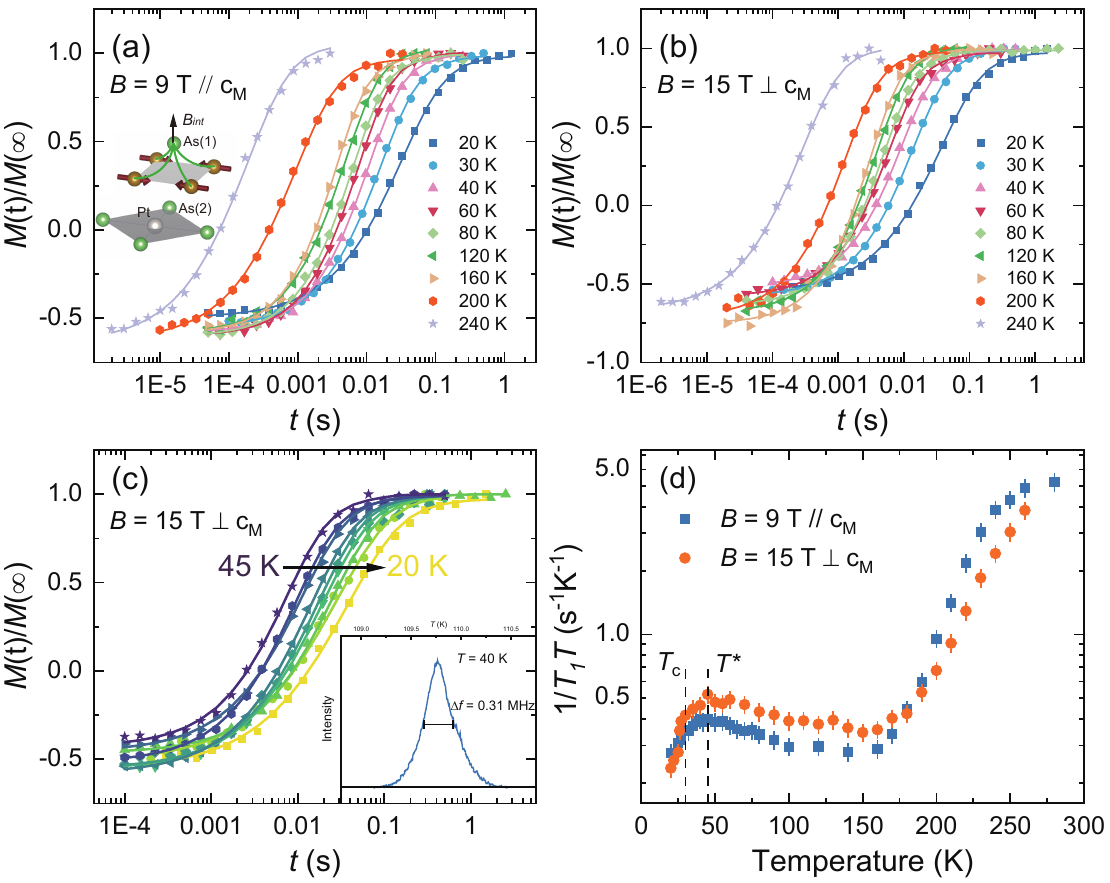}
	\caption{(Color online) NMR measurement results on CaLa-10-3-8 crystals.
		(a) Recovery of the nuclear magnetization as a function of delay time ($M(t)/M(\infty)$) at different temperatures, where the external magnetic field of 9 T is applied parallel to the $c_M$-axis ($B=9 $ T $\parallel c_M$).
        (b) $M(t)/M(\infty)$ results from $T=20$ K to 240 K for another measurement configuration $B=15 $ T $\perp c_M$.
        (c) Details of $M(t)/M(\infty)$ results from $T=20$ K to 45 K for the measurement configuration $B=15 $ T $\perp c_M$.
		(d) Temperature dependence of $1/T_1T$ obtained from the fitting results of magnetization recovery by Eq.1. The results are similar for two different measurement setup.
        The inset in (a) shows two inequivalent As sites and the internal hyperfine field at the As(1) site.
        The inset in (c) shows a typical $^{75}$As NMR spectrum measured with $B=15 $ T $\perp c_M$, where the peak width $\Delta f/f$ = 0.28$\%$ is smaller than previous reports in 10-3-8 system \cite{zhou2013, surmach2015}.
	}
\label{fig8}
\end{figure}

If there are preformed Cooper pairs above $T_c$ due to superconducting phase fluctuations, a partially opened pairing gap can be detected below $T^*$ in spectroscopic measurements \cite{emery1995}. In a Fermi liquid picture, the Knight shift ($K$) and the spin-lattice relaxation rate (1/$T_1$), deduced from the NMR spectrum, is sensitive to the DOS ($N(E_F)$) at the Fermi level. If we ignore the $q-$dependent spin fluctuations, both $K$ and 1/$T_1T$ follow a Korringa relation with $1/T_1T$ $\sim$ $K^2$ $\sim$ $N(E_F)^2$. In the underdoped cuprates, the loss of $N(E_F)$ by the pseudogap formation results in the decrease of $K$ and $1/T_1T$ far above $T_c$ \cite{timusk1999,norman2005}. We have performed NMR measurements at the $^{75}$As nuclear sites in CaLa-10-3-8 under two measurement configurations: $B=9 $ T $\parallel c_M$ and $B=15 $ T $\perp c_M$.  Fig. \ref{fig8}(a)-(c) show the recovery curves of the nuclear magnetization, and Fig. \ref{fig8}(d) shows the temperature dependence of the $1/T_1T$ obtained fitting of magnetization recovery curves in Fig. \ref{fig8}(a)-(c) . We don't obtain reliable $K$ for all measured temperatures since it is difficult to accurately determine the peak center in those broad resonance peaks due to the intrinsic disorders in the triclinic 10-3-8 system, even though the peak width of NMR spectrum $\Delta f/f$ = 0.28$\%$ is smaller than previous reports in 10-3-8 system (inset of Fig. \ref{fig8}(c)) \cite{zhou2013,surmach2015}.  Thus we only discuss the results of $1/T_1T$ obtained from the inversion-recovery method. As summarized in Fig. \ref{fig8}(d),  $1/T_1T$ firstly decreases rapidly at high temperatures, which has been extensively observed in other FeSCs such as BaFe$_2$As$_2$ and CaFe$_2$As$_2$. Then the $1/T_1T$ starts to increase as the antiferromagnetic spin coherence enhances the spin fluctuations, contrary to the constant $1/T_1T$ in a previous report on the CaPt-10-3-8 \cite{surmach2015}. Below $T^*$, the recovery curves return to the steady state more slowly (Fig. \ref{fig8}(c)), and then $1/T_1T$ starts to decrease again before a quick dropdown below $T_c$ (Fig. \ref{fig8}(d)), suggesting a reduction of $N(E_F)$, namely a gap opening on the Fermi surfaces. It should be noted that the temperature-dependent behavior of 1/$T_1T$ is almost the same for different directions and strengths of the magnetic field. The NMR results on CaLa-10-3-8 indicate that a pairing gap opens at $T^*$ instead of $T_c$, consistent with our results of INS and Nernst measurements.

\section{Discussions}
With accumulating evidence from transport, INS, Nernst, and NMR measurements, we have unambiguously observed the preformed Cooper pairs in the normal state of CaLa-10-3-8 superconductor. We notice that the pre-pairing temperature $T^*=45$ K for phase incoherence Cooper pairs is not very far above the zero-resistance temperature $T_{c0}=30$ K in CaLa-10-3-8 compound, which could be attributed to the smaller superconducting anisotropy and higher superfluid density in comparison with the layered cuprates. It seems that $\Delta T=(T^*-T_{c0}) < 20$ K in organic ion-intercalated FeSe superconductors ($T^*=60$ K and $T_{c0}=42$ K for (CTA)$_x$FeSe, $T^*=60$ K and $T_{c0}=43$ K for (TBA)$_x$FeSe, respectively) \cite{kang20201,Kang20222}. The gap forming temperature determined from photoemission measurements is either 83 K or $60 \sim 64$ K in the single-layer FeSe film, much higher than $T_{c0}= 23 \sim 30$ K obtained from electrical transport measurements \cite{whzhang2014,yxu2021,faeth2021}. However, we have to admit that it is kind of tricky to define the initial dropdown point of $1/T_1T$ above $T_c$ due to the crossover-like feature in its temperature dependence. For example, in the CsCa$_2$Fe$_4$As$_4$F$_2$ compound with similar $T_c=28.5$ K and $\mathit{\Gamma} \approx 14$, the reduction of $1/T_1T$ seems to happen below 90 K or 50 K, which depends on the criterion \cite{mcli2022}.  The scenario about incoherent Cooper pairing above $T_c$ has been put forward to understand the pseudogap state in cuprates \cite{emery1995}, as supported by numerous Nernst results \cite{xu2000,yayu2001,yayu2002,yayu2003,yayu2006}. But the situation is a bit of chaos, where the pre-pairing temperature $T^*$ defined by Nernst measurements may not coincide with the pseudogap temperature $T_p$ defined by the NMR and photoemission spectroscopies \cite{xu2000,yayu2001,yayu2002,yayu2003,yayu2006,ussishkin2002,kokanovic2009,chang2012,tafti2014,choiniere2018}. More controversies for the origin of pseudogap arise due to gap-like features from other competing orders in the underdoped region. In the iron pnictide superconductors, the Fe-As layer responsible for the superconductivity is relatively clean, and the Fe-Fe sublattice is not interrupted by the La doping in the 10-3-8 system, even there are some Pt disorders in Pt-As layer. Thus the definitions of $T^*$ obtained from various probes are more self-consistent than the case in cuprates. It should be noted that the possible inhomogeneous superconductivity cannot account for the pre-pairing feature at $T^*=45$ K, since the maximum $T_c$ at the optimal doping level is 33 K for the best condition in our CaLa-10-3-8 samples. Indeed, the large nonlinear field dependent Nernst signal, the BKT-like transition and the reduction of the density of state below $T^*$ all support the pre-pairing picture in CaLa-10-3-8 compound.

The precursor SRM and its field response are observed for the first time in FeSCs. Different from the lack of evidence of SRM in CaPt-10-3-8 and CaPt-10-4-8 compounds, the obvious spectral-weight redistribution and the appearance of a 5 meV spin gap in the superconducting state strongly support the existence of SRM in the CaLa-10-3-8, which is usually considered as the signature of $s^\pm$ superconducting pairing in between different Fermi pockets. In contrast to the other FeSCs, our INS results show no order-parameter like behavior in the temperature dependence of $\chi^{\prime\prime} (Q,\omega)$ at $E_R$, whereas the most rapid enhancement of its intensity occurs nearly at $T_c$. Instead, the slope of $\chi^{\prime\prime} (T)$ exhibits an anomaly at $T^* =$ 45 K associated with the onset of increasing magnetic response. This phenomenon is reminiscent of that shown in the underdoped cuprates such as YBa$_2$Cu$_3$O$_{6.6}$ and HgBa$_2$CuO$_{4+\delta}$. In cuprates, antiferromagnetic correlations have been argued to cause not only the $d-$wave superconductivity, but also the formation of pseudogap. The significant increase of the magnetic response below $T^*$ and the concomitant absence of a prominent effect across $T_c$ are taken as a prominent signature of the pseudogap state. Assuming the pseudogap has the same origin as the superconducting gap due to pairing electrons, by applying a magnetic field to suppress the pairing, the suppression of the SRM should persist above $T_c$ and terminate at $T^*$. This is exactly the case in our results on CaLa-10-3-8 (Fig. \ref{fig6} (f)). Therefore, the unusual temperature and field dependent behaviors of low-energy spin excitations in CaLa-10-3-8 may be ascribed to a pseudogap state, which probably is related to the spin-spin interactions. However, there are still some discrepancies with the INS results in cuprates. In the underdoped La$_{2-x}$Sr$_x$CuO$_4$ \cite{lee2003} and YBa$_2$Cu$_3$O$_{6+\delta}$ \cite{mignod1991,fong2000,pdai2001,mubhoff2021}, the low energy excitations below $E_R$ are partially suppressed even in the pseudogap state, forming a spin pseudogap. Nevertheless, a large full spin gap in the superconducting state is also observed in HgBa$_2$CuO$_{4+\delta}$ \cite{chan2016a,chan2016b}. Here in CaLa-10-3-8, the spin excitations only open a 5 meV gap in the superconducting state, which is likely due to a spectral weight redistribution of the SRM mode. The dispersions of spin excitations in the superconducting and pseudogap states are qualitatively different in YBa$_2$Cu$_3$O$_{6.6}$ ($T_c=61$ K, $T^*\approx 200$ K) and HgBa$_2$CuO$_{4+\delta}$ ($T_c=88$ K, $T^*=220$ K) with a clear SRM \cite{hinkov2007,chan2016b}, but keep almost the same Y-shape in HgBa$_2$CuO$_{4+\delta}$ ($T_c=71$ K, $T^*=305$ K) without a SRM \cite{chan2016a}. Although limited data in Fig. \ref{fig5}(a) suggest the low-energy dispersion in CaLa-10-3-8 may not change much below and above $T_c$, further time-of-flight neutron scattering experiments are highly desired to clarify this important issue, as well as the detailed dispersion of the SRM.

\section{Summary}

In summary, we have identified a precursor spin resonance mode in the normal state of CaLa-10-3-8 FeSC ($T_c=$ 30 K) with a characteristic temperature $T^{*}$= 45 K, corresponding to the onset temperature of the superconducting phase fluctuations determined by Nernst measurements and the reduction of DOS at the Fermi level determined by NMR measurements. More importantly, the magnetic field response of the resonance intensity persists above $T_c$ and terminates at $T^*$. Such phenomena indicate that the Cooper pairs can preform above $T_c$ in the iron pnictide superconductors, which closely resembles the case in the pseudogap state of underdoped cuprates. Our results enlighten the mechanism investigations on the pre-pairing picture and pseudogap phase in those quasi-2D high-$T_c$ superconductors.

\begin{acknowledgments}
This work is supported by the National Key Research and Development Program of China (Grant Nos. 2023YFA1406100, 2018YFA0704200 and 2020YFA0406000), the National Natural Science Foundation of China (Grant Nos. 11822411, 11874057 and 11961160699), the Strategic Priority Research Program (B) of the CAS (Grant Nos. XDB25000000 and XDB33000000) and K. C. Wong Education Foundation (GJTD-2020-01). H. L. is grateful for the support from the Youth Innovation Promotion Association of CAS (Grant No. Y202001). W. H. acknowledge the support from the Postdoctoral Innovative Talent program (Grant No. BX2021018) and the China Postdoctoral Science Foundation (Grant No. 2021M700250). This work is based on neutron scattering experiments performed at the Swiss Spallation Neutron Source (SINQ), Paul Scherrer Institut, Villigen, Switzerland (Proposal Nos. 20202223 and 20220955).
\end{acknowledgments}

\end{document}